\documentclass[11pt]{article}
\usepackage{amsmath,amsfonts,amssymb}
\usepackage[top=3cm, bottom=2cm, left=2cm, right=2cm]{geometry}
\usepackage{graphicx,subfigure}
\usepackage{pdfsync}

\newcommand{\be}{\begin{eqnarray}}
\newcommand{\ee}{\end{eqnarray}}
\newcommand{\beq}{\begin{eqnarray}}
\newcommand{\eeq}{\end{eqnarray}}
\newcommand{\bea}{\begin{eqnarray}}
\newcommand{\eea}{\end{eqnarray}}
\newcommand{\beas}{\begin{eqnarray*}}
\newcommand{\eeas}{\end{eqnarray*}}

\newcommand{\non}{\nonumber\\}
\newcommand{\bquo}{\begin{quote}}
\newcommand{\enqu}{\end{quote}}
\def\de{\partial}
\def\Tr{ \hbox{\rm Tr}}

\newcommand{\p}{\partial}

\def\bra{\langle}
\def\ket{\rangle}

\def\diag{\hbox{\rm diag}}

\def\NF{N_{f}}
\def\NC{N_{c}}
\def\sign{\hbox{\rm sign}}
\def\D{{\cal D}}
\begin{document}

\begin{titlepage}
{\hfill IFUP-TH/2011-13} 
\bigskip
\bigskip

\begin{center}
{\huge\bf Non-Abelian monopole-vortex complex }

\bigskip
\bigskip
{\large  
Mattia Cipriani ${}^{\dagger \, (1,3)}$, 
Daniele Dorigoni ${}^{\ddagger \, (2,3)}$, 
Sven Bjarke Gudnason ${}^{\star \, (4)}$, \\ 
Kenichi Konishi ${}^{\circ \, (1,3)}$ and 
Alberto Michelini $ ^{\diamond \, (2,3)}$
}

\bigskip
{\it  \footnotesize
${}^{(1)}$ Dipartimento di Fisica ``E. Fermi" -- Universit\`a di Pisa, 
 Largo Pontecorvo, 3, Ed. C, 56127 Pisa,  Italy \\
${}^{(2)}$ Scuola Normale Superiore - Pisa, 
  Piazza dei Cavalieri 7, Pisa, Italy \\
${}^{(3)}$ Istituto Nazionale di Fisica Nucleare -- Sezione di Pisa, 
  Largo Pontecorvo, 3, Ed. C, 56127 Pisa,  Italy \\
${}^{(4)}$ Racah Institute of Physics, The Hebrew University, 
  Jerusalem 91904, Israel
}

\vskip 2cm

{\bf Abstract}
\end {center}

In the context of softly broken ${\cal N}=2$ supersymmetric quantum
chromodynamics (SQCD), with a hierarchical gauge symmetry breaking 
$SU(N+1) \stackrel{v_{1}}{\longrightarrow} U(N) 
\stackrel{v_{2}}{\longrightarrow} \mathbf{1}$, $v_{1}\gg v_{2}$, we
construct monopole-vortex complex soliton-like solutions and examine
their properties. They represent the minimum of the static energy
under the constraint that the monopole and antimonopole positions
sitting at the extremes of the vortex are kept fixed. They
interpolate the 't Hooft-Polyakov-like regular monopole solution near
the monopole centers and a vortex solution far from them and in
between. The main result, obtained in the theory with $\NF=N$
equal-mass flavors, is concerned with the existence of exact
orientational $\mathbb{C}P^{N-1}$ zero modes, arising from the exact
color-flavor diagonal $SU(N)_{C+F}$ global symmetry. 
The ``unbroken'' subgroup $SU(N)\subset SU(N+1)$ with which the
na\"{i}ve notion of non-Abelian monopoles and the related
difficulties were associated, is explicitly broken at low
energies. The monopole transforms nevertheless according to the
fundamental representation of a new exact, unbroken $SU(N)$ symmetry
group, as does the vortex attached to it. We argue that this explains
the origin of the dual non-Abelian gauge symmetry. 

\vfill
\noindent
\rule{5cm}{0.5pt}\\
{\it\footnotesize 
$ \dagger $ cipriani(at)df.unipi.it \\ 
$ \ddagger$ d.dorigoni(at)sns.it  \\
$ \star $ gudnason(at)phys.huji.ac.il  \\
$ \circ $ konishi(at)df.unipi.it  \\
$ \diamond $ a.michelini(at)sns.it 
}

\end{titlepage}

\section{Introduction}

The last several years have witnessed a remarkable progress in our
understanding of vortex configurations in spontaneously broken gauge
theories, which carry continuous, non-Abelian internal zero modes: 
the non-Abelian vortices \cite{Hanany:2003hp,ABEKY}. Their rich
group-theoretical and dynamical properties have been subject of
intense study \cite{Konishi}-\cite{Eto:2010mu}. 

The physics of non-Abelian vortices is deeply related to the understanding
of the concept of the non-Abelian {\it monopole} and that of the quark
confinement, see e.g.~Ref.\cite{Konishi}. 
Indeed, a detailed, fully quantum-mechanical analysis of $4d$ gauge
theories with ${\cal N}=2$ supersymmetry has given important hints
about the low-energy, effective dual gauge symmetry. In particular,
fully quantum-mechanical light non-Abelian monopoles appear as the
infrared degrees of freedom in the so-called $r$ vacua of ${\cal N}=2$
supersymmetric QCD with $\NF$ quark multiplets, playing the role of
the order parameter for confinement (of non-Abelian variety) and for
dynamical symmetry breaking \cite{CKM}. 

The discovery of the non-Abelian vortex was partly motivated
\cite{ABEKY} by the desire to understand the physics of the quantum 
$r$-vacua, $r=2,\ldots, \NF/2$, from a more familiar semi-classical
viewpoint. In fact, semi-classically the connection between the
vortex solutions and regular 't Hooft-Polyakov monopoles arises from
the consideration of a hierarchical gauge symmetry breaking, e.g.
\beq 
SU(N+1) \stackrel{v_{1}}{\longrightarrow} U(N) 
\stackrel{v_{2}}{\longrightarrow} \mathbf{1} \ , \qquad 
v_{1} \gg v_{2} \gg \Lambda_{SU(N)}\ .   
\label{simplest}
\eeq
The monopole is supported by $\pi_{2}(SU(N+1)/U(N))=\mathbb{Z}$;
the low-energy vortex solutions correspond to non-trivial elements of
$\pi_{1}(U(N))=\mathbb{Z}$. The exact sequence of homotopy groups
relates the two solitons of different co-dimension \cite{Duality}, and
the global symmetry consideration tells us that the non-Abelian 
{\it vortex} implies non-Abelian {\it monopoles} sitting at its
extremes. For $\infty > v_{1}\gg v_{2}> 0$, one is inevitably led to
consider metastable monopole-vortex complex solitons. 

The aim of this paper is to pursue further the study of such a
monopole-vortex complex, including the numerical analysis of the field
configurations involving both the magnetic monopole region and the
asymptotic vortex-like region, with all fields approaching smoothly 
their vacuum expectation value (VEV) away from the complex. In this
sense this paper is a continuation of the work by Auzzi
et.~al.~\cite{ABEK}. 
We clarify also some aspects of the non-Abelian orientational moduli, 
extensively studied in the last several years in the context of the
vortex solutions, and show how the properties of the non-Abelian
orientational moduli can be extended to the whole monopole-vortex
complex.

The organization of the paper will be the following.
In Sec. 2 after a brief introduction of the model we will focus our attention
to the main characters of our paper: the vortex and the monopole, a particular care will be given to the study of the symmetries 
present at different energy scales.
In Sec. 3 these two objects will be glued together and we will study the monopole-vortex complex as a whole.
We will present both numerical and analytical results concerning the profile functions for the various fields and
for the magnetic field, furthermore through a careful identification of the symmetries
possessed by the complex we will be able to analyze the low energy dynamics of the orientational zero-modes living on the complex.
The conclusions and results obtained in this work are presented in Sec. 4 while we give a more detailed analysis 
of the full set of equations of motion and the consistency of our Ansatz in the two appendices.

\section{The model}

To be concrete, we shall work with the softly broken ${\cal N}=2$
supersymmetric models, due to the many advantages they offer. The
fields are those of the ${\cal N}=2$ gauge multiplet (the gauge field,
the gauge fermion, the adjoint scalar and fermion) together with
hypermultiplets. To be precise, we take an $SU(N_{c})$  ($N_{c}=N+1$) gauge theory
with $\NF=N$ flavors of hypermultiplets (quarks), and the mass
parameters are tuned such that at two hierarchically different scales 
the gauge symmetry is broken as in Eq.~(\ref{simplest}). 
The Lagrangian of the underlying $SU(N+1)$ theory has the structure
\begin{align}
\mathcal{L} &= \frac{1}{4\pi} {\rm Im} \, \tau 
\left[\int d^4 \theta \,
\Tr \left(\Phi^\dag e^{-2V} \Phi\right) 
+ \frac{1}{2} \int d^2 \theta\, \Tr \left(W^\alpha W_\alpha\right) \right]
+ \mathcal{L}^{\rm(quarks)} 
+ \int d^2\theta \,\mu \,\Tr \,\Phi^2 \ , 
\label{lagrangian} \\
\mathcal{L}^{\rm(quarks)} &= 
\sum_i \left[ \int d^4 \theta \; 
\left(Q_i^\dag e^{-2V} Q^i 
+ \tilde{Q}_i e^{2V} \tilde{Q}^{{\dag}i} \right) 
+ \int d^2 \theta \, \left( \sqrt{2} \, {\tilde Q}_i \Phi Q^i +
m_{i} \,  {\tilde 
  Q}_i Q^i   \right) \right] \ ,
\label{lagquarkN}
\end{align}
where $m_{i}$ are the bare masses of the quark fields and the complex
coupling constant reads
\beq
\tau \equiv \frac{\theta}{2\pi} + \frac{4\pi i}{g^2} \ .  
\label{struc}
\eeq   
The mass of the adjoint chiral multiplet $\mu$ breaks supersymmetry to
${\cal N}=1$. 

After elimination of the auxiliary fields the bosonic Lagrangian
takes the form
\begin{align}
\mathcal{L} &= 
-\frac{1}{4g^2} \left(F_{\mu\nu}^{A}\right)^2 
+\frac{1}{g^2} |{\cal D}_{\mu} \phi^A|^2 
+\left|{\cal D}_{\mu} q^i\right|^2 
+\big|{\cal D}_{\mu} \tilde{q}^{\dag i}\big|^2 
- V_1 - V_2 \ , 
\label{Lag} \\
V_1 &= \frac{1}{8} \sum_A 
\left[-\frac{i}{g^2}f^{ABC} \phi^{B \dag} \phi^C 
+ q_i^\dag t^A q^i 
- \tilde{q}_i t^A \tilde{q}^{\dag i} \right]^2
= \frac{1}{8}\sum_A 
\left((T^A)_a^b \,  \left[ -\frac{2}{g^2}\, [\phi^\dag, \phi]_b^a  
  + q_i^{\dag a} q_b^i 
  - \tilde{q}_i^a \tilde{q}_b^{\dag i}
  \right]\right)^2 , \nonumber \\ 
V_2 &= 
g^2 \sum_A \left|\mu \, \phi^A +\sqrt{2} \, \tilde{q}_i\, t^A q^i \right|^2  
+ \left|[m_i+\sqrt{2}\phi]^\dag \, \tilde{q}^{\dag i}\right|^2  
+ \left|[m_i+\sqrt{2}\phi]  \, q^i\right|^2 \ ,   \label{nonbps} 
\end{align}
where $A=1,2,\ldots,(N+1)^{2}-1$ and the sum over $i=1,2,\ldots,\NF$
as well as $a,b=1,2,\ldots,N+1$ is implicit.
In the construction of the monopole-vortex complex soliton solutions
it turns out to be sufficient to consider the VEVs and fluctuations
around them which satisfy 
\beq   
[\phi^\dag, \phi] = 0 \ , \qquad 
q^i = \tilde{q}^{\dag i} \ ,
\eeq
therefore $V_1$ can be set identically to zero in what follows.

The vacuum expectation values (VEVs) of the scalar fields are
determined from the minima of the potential following from
Eq.~(\ref{nonbps}), e.g.~see Ref.~\cite{CKM}. They are found to be 
\begin{align}
q_a^i &= \delta_{a}^{i}\, d_{i} \ , \quad 
\tilde{q}_i^a = \delta_i^a \, \tilde{d}_i \ , \qquad\,
{\rm for } \ \ 
i=1,2,\dots,r, \quad 
a=1,2,\ldots, N+1 \ ; \\
q_a^i &= 0 \ , \quad
\tilde{q}_i^a = 0 \ , \qquad\qquad\quad
{\rm for } \ \ 
i=r+1,\ldots, \NF \ ; \label{vevofq} \\
d_i \tilde{d}_i &= \mu\, m_i 
+ \frac{\mu}{N+1-r} \sum_{k=1}^r m_k \ , \qquad  
(d_{i} > 0) \ , \qquad  
|\tilde{d}_i| = d_i\ , \label{solnd} \\
\Phi &= \frac{1}{\sqrt{2}}\diag \, (-m_1, -m_2, \ldots, -m_r, c,
\ldots , c) \ ; \qquad 
c = \frac{1}{N+1-r} \sum_{k=1}^r m_k \ , 
\label{diagphi}
\end{align}
where the integer 
\beq 
r = 0,1, \ldots, \min\,\{\NF,\NC-1\},\qquad \NC=N+1 \ ,
\eeq
labels the possible (classical) vacua. The vacua of a given $r$ are
further classified according to which set of $r$ (out of $\NF$) masses
are used to construct the solution, leading to the total of 
\beq
\#{\rm vacua} = \sum_{r=0}^{\min\,\{\NF,\NC-1\}}\, (\NC-r) \, 
\begin{pmatrix} \NF \\ r \end{pmatrix} \ .
\label{nofvac}
\eeq

As explained in Ref.~\cite{CKM}, by choosing a generic set of bare
masses $m_{i}$ and by deforming with the ${\cal N}=1$ mass term
$\mu\,\Tr\,\Phi^2$ the continuous vacuum degeneracy is lifted
altogether, leaving this discrete set of vacua. At small $m_{i}$
and $\mu$ ($\ll \Lambda$) the interactions become strong in the
infrared, in all these vacua. These are indeed the vacua we are
interested in.\footnote{This is one of the motivations for considering
  the system with generic masses and with $\mu \ne 0$ first, and then
  eventually taking the equal-mass or massless limit. On the
  contrary, if we considered directly the massless theory, or equal
  mass cases, we would find flat directions (continuum vacuum
  degeneracy); at a generic point along such Higgs branches, the
  coupling constant remain small at all scales. } 

By tuning the bare squark masses $m_{i}$ to an equal, common value
$m$, we see that an exact color-flavor diagonal $SU(r)$ symmetry
survives in a vacuum with a given $r$. 
\emph{For definiteness, below we shall work with the classical 
$r=N$ vacuum  where} 
\beq 
\bra \phi \ket = v_{1} 
\begin{pmatrix}{\bf 1}_{N\times N} & {\bf 0}_{N \times 1} \\
{\bf 0}_{1 \times N}  & -N
\end{pmatrix} \ , \qquad 
\bra q \ket = \bra \tilde{q}^\dag \ket = v_{2}
\begin{pmatrix}
{\bf 1}_{N\times N} \\
{\bf 0}_{1 \times N}
\end{pmatrix} \ . \label{VEVS}
\eeq
and
\beq 
v_{1} \equiv -\frac{m}{\sqrt{2}} \ , \qquad
v_{2} \equiv \sqrt{(N+1)\, m \,\mu} \ . 
\eeq
are obtained by taking such a limit.
Note that in this vacuum the \emph{gauge group} $SU(N+1)$ is
completely broken; at the same time, however, the color-flavor
diagonal {\it global} $SU(N)$ symmetry  
\beq 
q^{U} = 
\begin{pmatrix} 
U & \\ & 1
\end{pmatrix}\, q \, U^{-1} \ , \qquad 
\left(\phi^{U}, A_{i}^{U}\right) = 
\begin{pmatrix}
U & \\ 
& 1\end{pmatrix} \, 
\left(\phi, A_{i} \right) \, 
\begin{pmatrix}
U^{-1} &  \\
& 1\end{pmatrix} \ , \label{colorflavor}
\eeq
($U \in SU(N)\subset SU(N+1)$) remains unbroken by both VEVs. It is an
exact global symmetry of the whole system. The system is in the
so-called color-flavor locked phase. A hierarchical symmetry breaking
pattern (\ref{simplest}) is realized if 
\beq 
|m| \gg |\mu| \gg \Lambda \ , \qquad 
\therefore \qquad 
|v_{1}| \gg  |v_{2}| \ . 
\label{hierarchy} 
\eeq

\subsection*{Remarks} 

\begin{description}

\item[(i)] The terms containing the adjoint scalar mass $\mu$ play
  two crucial roles in our model. On the one hand, they induce the
  small squark condensates, (\ref{VEVS}), bringing the system into
  a completely Higgsed phase. The existence of the vortex solutions in
  the low-energy approximation and their properties, all rely on this
  parameter. Note that due to supersymmetry, the high-energy
  approximate monopole solution ($v_{2}=0$) and low-energy 
  approximate vortex solutions  ($v_{1}=\infty$) are both
  BPS-saturated. 

On the other hand, non-vanishing $\mu$ introduces terms in $V_2$ which
make both the low-energy vortex and high-energy monopole ``solutions'' 
unstable (non-BPS). It is these terms which allow the two solitons of
different codimensions to get smoothly combined into a monopole-vortex 
complex.  

\item[(ii)] Of course such a complex ``soliton'' is not a true
  solution of the field's equations of motion; it is only so under the 
  condition that the monopole center positions are kept fixed. Under
  the assumption of a hierarchical gauge symmetry breaking
  (\ref{hierarchy}), this is not a problem: it is a perfectly sensible
  (Born-Oppenheimer) procedure, as the motion of the massive
  monopole can be neglected in the first instance, in the study of
  low-energy fluctuations of {\it  orientational zero modes} of the
  whole complex.  

\item[(iii)]  Actually the case for working with non-BPS objects as
  these complex solitons can be made stronger. Just as in the case of
  the real-world quark-antiquark-chromoelectric string composites
  (the mesons), our monopole-vortex-antimonopole complex will get
  stabilized after the quantization of the radial or rotational
  motions are appropriately taken into account. Of course, only the
  ground state for each flavor quantum number will be truly stable,
  as the pair production of the monopole-antimonopole from the vacuum,
  though suppressed, cannot be rigorously set to zero. But the same
  holds for the real-world meson states! The fact that these are not
  topologically stable {\it as static configurations}, is therefore
  not a shortcoming at all; rather it is the necessary price to pay  
  to understand the real-world confinement mechanism. 

\item[(iv)]  Our semi-classical consideration is valid for the
  parameter region 
  \[ |m_{i}|=|m| \gg |\mu| \gg \Lambda\ .  \]
  As the bare quark masses and the adjoint scalar mass $\mu$ are
  decreased to small values of the order of $\Lambda$, the system
  becomes strongly coupled and the classical $r=N$ vacua we started
  with turn into the weakly coupled, dual quantum $r=0$ vacua
  \cite{CKM}. This means that the vortex (and monopole) orientational
  zero modes fluctuate strongly and Abelianize. A possible route to
  reach the quantum $r$ vacua ($2 \le r \le \NF/2$), where light
  non-Abelian monopoles appear as the infrared degrees of freedom,
  through a careful tuning of bare masses and through the consequent
  vortex solutions with orientational moduli living on
  $\mathbb{C}P^{r-1}\times \mathbb{C}P^{N-r-1}$, was discussed by some
  of us in Ref.~\cite{DKO}.

\end{description}

\subsection{Vortex solution far from the monopole center}

In the vacuum (\ref{VEVS}) the $(N+1)$-th color component of the squark
fields has a mass of the order of $v_{1}$ and can be integrated away,
in the study the low-energy dynamics. 
At mass scales much lower than $v_{1}$ the theory is an 
$SU(N)\times U(1)$ gauge theory with $\NF=N$ massless flavors, in the
color-flavor locked phase. 
The vortex solutions in these systems have been a subject of an
intense study for the last several years
\cite{Konishi}-\cite{Eto:2010mu}. A lowest-winding vortex solution
oriented in the $(1,1)$ direction in color-flavor space\footnote{These 
  vortices are precisely those studied by \cite{AEV} for the special
  case of $N=2$. The $(N+1)$-th color component of the (massive) squark
  fields is vanishing and not shown in Eq.~(\ref{LEvortexSU2}).} has the
form 
\begin{align}
q  &= 
\begin{pmatrix}
e^{i\varphi}q_1(\rho) & \\
& q_2(\rho) \mathbf{1}_{N-1} 
\end{pmatrix}\ ;   \non
A_{i} &= \epsilon_{ij}\frac{x^j}{\rho^2}\left[
\frac{1-f(\rho)}{N}
\begin{pmatrix}
1 \\
& \mathbf{1}_{N-1} & \\
& & -N
\end{pmatrix}
+ \frac{1-f_{\rm NA}(\rho)}{N} 
\begin{pmatrix}
N-1 &  & \\
& -\mathbf{1}_{N-1} &  \\
& & 0
\end{pmatrix}\right] \ , \quad i=1,2 \ , \non
\phi &= \left(v_{1} + \frac{\lambda(\rho)}{\sqrt{2N(N+1)}}\right)
\begin{pmatrix}
1 \\
& \mathbf{1}_{N-1} & \\
& & -N
\end{pmatrix} 
+ \frac{\lambda_{\rm NA}(\rho)}{\sqrt{2N(N-1)}}
\begin{pmatrix}
N-1 &  & \\
& -\mathbf{1}_{N-1} &  \\
& & 0
\end{pmatrix} \ , \label{LEvortexSU2}
\end{align}
with the profile functions satisfying the appropriate boundary
conditions 
\begin{align}
&q_{1,2}(\infty) = v_{2} \ , \quad
f(\infty) = f_{\rm NA}(\infty) = \lambda(\infty) 
  = \lambda_{\rm NA}(\infty) = 0 \ , \non
&q_1(0) = 0 \ , \quad\quad\;\ 
\p_\rho \, q_2(0) = \p_\rho\, \lambda(0) = \p_\rho\, \lambda_{\rm NA}(0) = 0
\ , \quad  
f(0) = f_{\rm NA}(0) = 1 \ . \label{boundaryc}
\end{align}
We have introduced above the cylindrical coordinates 
\[  z\ ,\quad  \rho=\sqrt{x^{2}+ y^{2}} \ , \quad  
  \varphi = \tan^{-1}\left(\frac{y}{x}\right) \ , \]
which are convenient in the description of the vortex configuration;
the gauge field above in fact contains only the $\varphi$ component,
$A_{\varphi} = -\frac{1}{\rho} [\cdots]$ where $[\cdots] $ is the
expression in the square bracket appearing in Eq.~(\ref{LEvortexSU2}).

The $\mu$ dependent terms in Eq. (\ref{nonbps}), coming from the
adjoint scalar mass term $\mu\Tr\Phi^2$, imply that the field $\phi$ 
behaves non-trivially (Eq.~(\ref{LEvortexSU2})): the vortex solution
is necessarily non-BPS \cite{AEV}. Its properties have been
carefully studied by Auzzi et.~al.~\cite{AEV} in the case of
the low-energy $SU(2)\times U(1)$, $\NF=2$, theory. It has been shown 
that, independent of the sign of $\mu, $ or of $m_{1}=m_{2}$, the
vortex tension is less than the BPS vortex (for $\mu=0$). The behavior
of the profile functions is quite similar, for small $\mu$, to the
case of the BPS vortex analyzed earlier, except for the presence of
the small non-vanishing profile functions for $\phi$ in and around the 
vortex core.

For our purpose below, it is necessary to consider these vortex
solutions in a gauge where the scalar quark fields do not wind at
infinity. The $SU(N)\times U(1) \subset SU(N+1)$ gauge
transformation needed is 
\beq 
U^{\rm (singular)} = 
\begin{pmatrix}
e^{-i \varphi} &   &  \\ 
& \mathbf{1}_{N-1} & \\
& & e^{i \varphi}
\end{pmatrix} \ , 
\eeq
the vortex solution is indeed transformed into the form
\begin{align}
q &= 
\begin{pmatrix}
q_1(\rho) & \\
& q_2(\rho) \mathbf{1}_{N-1}
\end{pmatrix} \ ;   \label{qsing} \\
A_{\varphi}  &= \frac{1}{\rho}  \left[
\frac{f(\rho)}{N}
\begin{pmatrix}
1 \\
& \mathbf{1}_{N-1} & \\ 
& & -N
\end{pmatrix}
+\frac{f_{\rm NA}(\rho)}{N} 
\begin{pmatrix}
N-1 &  & \\ 
& -{\mathbf 1}_{N-1} &  \\
& & 0
\end{pmatrix} \right] \ , \label{LEvortexSing}
\end{align} 
while the $\phi$ field remains invariant as in Eq.~(\ref{LEvortex}). 
In this (singular) gauge all the topological features are hidden into
the gauge-field singularity along the vortex core ($\rho=0$),  
\beq 
A_{\varphi}(\rho) \ \stackrel{\rho \to 0}{\longrightarrow} \ 
\frac{1}{\rho}
\begin{pmatrix}
1 & 0 & 0 \\
0 & {\bf 0}_{N-1} & 0 \\
0 & 0 & -1
\end{pmatrix} \ . \label{Dirac1}
\eeq
The magnetic flux is nevertheless regular everywhere and given by 
\beq
B_{z} = \frac{1}{\rho} \frac{\de(\rho A_{\varphi})}{\de\rho} 
= \frac{1}{\rho} \left[
\frac{\de f(\rho)}{\de\rho}\frac{1}{N}
\begin{pmatrix}
1 \\
& \mathbf{1}_{N-1}  \\
&  & -N
\end{pmatrix}
+\frac{\de f_{\rm NA}(\rho) }{\de \rho } \frac{1}{N} 
\begin{pmatrix}
N-1 & & \\
& -\mathbf{1}_{N-1} & \\
&  & 0\end{pmatrix} \right] \ , \label{LEvortex}
\eeq
with the $B_{x},B_{y}$ components vanishing, so that the
total flux is given by 
\beq
\mathcal{F}_{z} = 2\pi \int d\rho \,\rho\, B_{z} 
= - 2\pi \, 
\begin{pmatrix}
1 & 0 & 0 \\
0 & {\mathbf 0}_{N-1} & 0 \\
0 & 0 & -1
\end{pmatrix} \ . \label{VortexFlux}
\eeq
Note that such a vortex leaves an $SU(N-1) \times U(1)$ subgroup of
the color-flavor diagonal global symmetry $SU(N)_{C+F}$ intact, and as
a result develops orientational zero modes living in 
\beq
\mathbb{C}P^{N-1} \sim \frac{SU(N)}{SU(N-1) \times U(1)} \ ; \label{coset}
\eeq
they are sort of Nambu-Goldstone modes, propagating along the vortex
string. 
More concretely, there is a continuous infinity of degenerate vortex 
solutions related by
\beq
q^{U} = 
\begin{pmatrix}
U & \\
 & 1
\end{pmatrix} \, q \, U^{-1} \ , \quad
A_{i}^{U} =
\begin{pmatrix}
U & \\
& 1
\end{pmatrix} \, A_{i} \,
\begin{pmatrix}
U^{-1} &  \\
& 1
\end{pmatrix} \ , \quad
\phi^{U} =
\begin{pmatrix}
U & \\
& 1
\end{pmatrix} \, \phi \, 
\begin{pmatrix}
U^{-1} & \\
& 1
\end{pmatrix} \ . \label{SUNRotation}
\eeq
Note that the color-flavor $N \times N$ $SU(N)$ matrix $U$ belongs to
the coset Eq.~(\ref{coset}) as an individual vortex solution
(\ref{LEvortexSing}) preserves the subgroup $SU(N-1) \times U(1)$
generated by 
\beq
\begin{pmatrix}
0 & 0 \\
0 & e^{i\alpha_A T^A}
\end{pmatrix} \subset U \ , \qquad
\begin{pmatrix}
e^{i(N-1)\alpha_0} & 0 \\
0 & e^{-i\alpha_0}\mathbf{1}_{N-1}
\end{pmatrix} \subset U \ , \label{invariant}
\eeq
where the $T^{A}$'s are the standard $SU(N-1)$ generators. 

The existence of the exact orientational zero modes of
Eq.~(\ref{SUNRotation}) possessed by the low-energy vortex solutions
characterizes these solitons as \emph{non-Abelian vortices}
\cite{Hanany:2003hp,ABEKY}. They possess a genuine moduli space of
solutions, all having the same tension. 
Their low-energy fluctuations (i.e.~fluctuations carrying energies
much lower than the characteristic vortex scale, $v_{2}$) can be shown
to be effectively described as a two-dimensional sigma model on
$\mathbb{C}P^{N-1} = SU(N)/SU(N-1)\times U(1)$.  Dynamical features of
these fluctuations, detailed moduli-space structures, and their
group-theoretic properties have been the focus of considerable
attention in the last several years \cite{Konishi}-\cite{Eto:2010mu}.

\subsection*{Remarks}
\begin{description}
\item[(i)]  Note that even though the light squark fields have only
  the first $N$ color components, the gauge field has a non-vanishing
  $(N+1, N+1)$ element (see Eq.~(\ref{LEvortex}) or
  Eq.~(\ref{LEvortexSing})) due to the fact that the $U(1)$ gauge
  group descends from the underlying $SU(N+1)$ gauge group. This turns
  out to be crucial in the consideration of the vortex-monopole
  complex below.  
  In this respect, the present model (i.e.~the model of
  \cite{ABEKY,ABEK,AEV}), where the $SU(N)\times U(1)$ theory arises
  as a low-energy approximation of a spontaneously broken $SU(N+1)$
  theory, differs essentially from the genuine $U(N)$ model studied by
  other groups
  \cite{Hanany:2003hp},\cite{Hanany:2004ea}-\cite{Eto:2005yh}, where 
  all scalar and gauge field components live in $N\times N$ color
  space.
  Even though they share many interesting features of the non-Abelian
  vortex solutions with our model, it is not possible in these latter
  models to relate the non-Abelian orientational moduli of the
  low-energy vortex to the notion of the non-Abelian monopole, which
  ``lives'' in the larger $SU(N+1)$ gauge group space.
\item [(ii)] When the $SU(N)$ and $U(1)$ coupling constants are set to
  be equal -- this would be the case for the theory just below the
  higher gauge-symmetry breaking mass scale $v_{1}$ -- the vortex
  solution above reduces exactly to the Abelian ANO vortex \cite{AEV},
  \emph{embedded} in the $(1,1)$ corner of the color-flavor space.  We
  shall use below such a simplification e.g.~for solving the
  monopole-vortex complex numerically (Subsection~\ref{Numerical}), 
  but the more general discussion on the orientational moduli
  (Subsection~\ref{Orientation}) of the complex is independent of it.

\end{description}

\subsection{The monopole}

As the underlying $SU(N+1)$ gauge group is simply connected, the
vortex solutions reviewed above cannot be stable in the full theory,
which contains the massive monopole excitations.  A vortex must end at
the two endpoints where monopoles related to the symmetry breaking
$SU(N+1)\to SU(N)\times U(1)$ are situated. The properties of such a
monopole solution can be studied to first approximation by neglecting
the small VEV, $v_{2}$, or equivalently, by studying the system
sufficiently close to the monopole center, i.e.~at the distances $R$ 
\beq 
\frac{1}{|v_{2}|} \gg  R \sim  \frac{1}{|v_{1}|} \ , \label{HEapprox}
\eeq
from the center. For $q=\tilde{q}^\dag=0$ and for real $\phi^A$ neither 
$V_1$ nor $V_2$ contribute, so the field configuration
inside a sphere of such radius $R$ should be approximately equal to
the standard 't Hooft-Polyakov BPS monopole solution \cite{TH},
embedded in an appropriate corner of $SU(N+1)$ gauge group. By
choosing an $SU(2)\subset SU(N+1)$ group generated by
\beq
S_1 = \frac{1}{2}
\begin{pmatrix}
0 & & 1 \\
& {\bf 0}_{N-1} & \\
1 & & 0\end{pmatrix} \ , \quad 
S_2 = \frac{1}{2}
\begin{pmatrix}
0 & & -i \\
& {\bf 0}_{N-1} & \\
i & & 0
\end{pmatrix} \ , \quad 
S_3 = \frac{1}{2}
\begin{pmatrix}
1 & & 0 \\
& {\bf 0}_{N-1} & \\
0 & & -1
\end{pmatrix} \ , \label{SU2}
\eeq 
(which is broken to $U(1)$ by the VEV of $\phi$) a monopole solution
can be constructed explicitly as \cite{EW}: 
\beq
A_i({\bf r}) = A_i^a({\bf r}) \, S_a; \qquad 
\phi({\bf r}) = (N+1) \, v_{1} 
\frac{r^{a} \, S_a}{r}\chi(r) 
+ v_{1} 
\begin{pmatrix} 
- \frac{N-1}{2} & & \\
& \mathbf{1}_{N-1} & \\
& & -\frac{N-1}{2}
\end{pmatrix} \ , 
\eeq
where\footnote{The index $a=1,2,3$ refers to the $SU(2)$ group
  utilized to construct the solution; the gauge field $A_{i}$ and the
  adjoint scalar $\phi$ are both $(N+1)\times (N+1)$ matrices in the
  $SU(N+1)$ color group. } 
\beq
A_i^a({\bf r}) = \epsilon_{aji} \frac{r^j}{r^2} a(r) \ ,
\eeq
is the standard 't Hooft-Polyakov-BPS solution with 
\beq
a(r) = 1 - \frac{g v_{1} r}{\sinh(g v_{1} r)} \ ;  \qquad
\chi(r) = \coth(g v_{1} r) - \frac{1}{g v_{1} r} \ ,  
\eeq
the latter behaving asymptotically as 
\beq
a(r) \to 1 \ , \qquad 
\chi(r) \to \sign (v_{1}) \ , \qquad 
r \to \infty \ , 
\eeq
(for the antimonopole, $\chi(r) \to -\chi (r)$).
The constant in the $\phi({\bf r})$ field is added so that it reduces 
asymptotically to the vacuum expectation value of Eq.~(\ref{VEVS}), in
a fixed (here chosen as $(0,0,\infty)$) direction. The ``magnetic
flux'' emanated from the magnetic monopole is given by   
\beq
B_{i} = \frac{1}{2}\epsilon_{ijk} F_{jk} \ \ 
{\stackrel {r \to \infty} {\longrightarrow}} \ \ 
\frac{r_{i}({\bf S}\cdot {\bf r})}{r^{4}} \ ,  \label{Magflmono}
\eeq 
which of course is a well-known radial Dirac monopole field, embedded 
in the $S_{i}$ color directions.

The ``monopole'' mass (the energy of the configuration around its
center) can be approximately calculated as\footnote{We take the adjoint scalar
  field $\phi$ to be real here and hence normalize it canonically as a
  real scalar field. } 
\beq
H = \int_{|{\bf r}|<R} d^3x \, \Tr \, \left[
\frac{1}{2 g^2}(F_{ij})^2 
+\frac{1}{g^2}|{\cal D}_{i} \phi|^2 \right] \ .
\eeq
Rewriting the Hamiltonian as
\beq 
H = \frac{1}{g^{2}} \int_{|{\bf r}|<R} d^3x  \, \Tr \, \left[
\frac{1}{2} |F_{ij} - \epsilon_{ijk}({\cal D}_{k}\phi)|^2
+\de_k (\epsilon_{ijk}F_{ij}\,\phi)\right]
\eeq
our monopole configuration is seen to satisfy approximately the
non-Abelian Bogomol'nyi equations 
\beq
B_k = {\cal D}_{k} \phi \ , \label{nbe}
\eeq
so that the monopole mass is given approximately by 
\beq
H = \frac{2}{g^{2}} \int_{|{\bf r}|=R} d{\bf S}\cdot \Tr \, 
  [\phi \, {\bf B}]
  = \frac{4\pi}{g^{2}} \, (N+1) \, |v_1| \ ,
\eeq
where $R$ is in the range (\ref{HEapprox}).
The monopole mass is, naturally, gauge independent. On the contrary,
the ``magnetic flux'' (\ref{Magflmono}) is a gauge dependent
quantity.  To match the ``vortex part'' of the solution discussed in the previous subsection,  
it is necessary to choose a particular gauge in which the adjoint
scalar field does not wind at infinity (the so-called singular gauge)
in order to describe the complex.  
By introducing the spherical coordinates
\[ \frac{{\bf r}}{r} = 
(\sin\theta \cos\varphi, \sin\theta \sin\varphi, \cos\theta) \ , 
\]
the required gauge transformation is given by 
\beq
\phi \to V\, \phi \, V^\dag \ , \quad
A_{i} \to  V \left(A_{i} - i\de_{i}\right) V^\dag \ ,
\qquad
V =  e^{-i \varphi  S_{3}} e^{i \theta S_{2}} e^{i \varphi S_{3}} \ ,   
\eeq
where the $S_{i}$'s are the $SU(2)\subset SU(N+1)$ generators of
Eq.~(\ref{SU2}). 
The adjoint scalar field is simply transformed to the fixed direction
in the color space (chosen here in the $S_{3}$ direction): 
\beq 
\phi({\bf r}) = (N+1) \, v_{1} \, S_3 \, \chi(r)
+ v_{1}
\begin{pmatrix}
-\frac{N-1}{2} &  & \\
& {\bf 1}_{N-1} & \\
& & -\frac{N-1}{2}
\end{pmatrix} \ , \label{NAmonop}  
\eeq
whereas the gauge field has a slightly more complicated form.  In view
of our task below, of connecting the monopole solution to the
low-energy vortex, it is convenient to express the gauge field in the
new gauge in the components in the cylindrical coordinate system:  
\beq  A_{\rho}   &=&  \frac{\cos\theta}{r} 
(S_{2} \cos\varphi - S_{1} \sin\varphi)(a(r) - 1) \ ; \non
A_{\varphi} &=& \frac{1}{\rho} \left[ S_{3}(1 - \cos\theta) 
- \sin\theta (S_{1} \cos\varphi + S_{2} \sin\varphi)(a(r) - 1)
\right]\ ; \label{Manogauge} \\
A_{z} &=& - \frac{\sin\theta}{r} 
(S_{2} \cos\varphi - S_{1} \sin\varphi)(a(r) - 1) \ , \nonumber 
\eeq
where the conversion between spherical and cylindrical coordinates
involve the relation 
\beq 
z = r\, \cos\theta \ , \qquad 
\rho = r\, \sin\theta \ , 
\eeq
while the azimuthal angle $\varphi$ is common in the two coordinate
systems.   
The gauge field displays the well-known Dirac string singularity along
the negative $z$ direction in this gauge: 
\beq
\left.A_{\varphi}\right|_{\theta=\pi} \sim \frac{2 S_{3}}{\rho} \ . 
\label{Dirac2}
\eeq

The asymptotic ($r \gg \frac{1}{|v_{1}|}$) behavior of the magnetic
field ${\bf B}$ can be found by dropping terms multiplied by the
factor $(a(r) - 1)$ appearing in Eq.~(\ref{Manogauge}): only
$A_{\varphi}$ survives and  
\beq
B_{\rho} \simeq S_{3} \frac{\rho}{r^3} \ , \qquad
B_{\varphi} \simeq 0 \ , \qquad
B_{z} \simeq S_{3} \frac{z}{r^3} \ ,
\eeq
which of course is a well-known radial Dirac monopole field, embedded
in the $S_{3}$ color direction. This radial magnetic field can be seen
more easily by directly transforming the asymptotic magnetic field in
Eq.~(\ref{Magflmono}) by $B_i \to V B_{i} V^\dag$,
which yields
\beq 
B_{i} \simeq  \frac{r_{i} S_{3}}{r^{3}}\ ,    
\eeq
The total magnetic flux through the surface of a sphere of radius $R$
around the center, is then given by  
\beq 
\int_{|{\bf r}|=R} d{\bf S}\cdot {\bf B} = 4 \pi S_{3} = 2\pi
\begin{pmatrix}
1 &  & 0 \\
& {\bf 0}_{N-1} &  \\
0 &  & -1
\end{pmatrix} \ , \label{Monoflux}
\eeq
independently of $R$.

\section{Monopole-vortex complex} 

The equality in the magnitude between the vortex flux
(\ref{VortexFlux}) through a plane perpendicular to the vortex axis,
and the magnetic monopole flux (\ref{Monoflux}) through the surface of
a tiny sphere around the monopole, is an example of the ``flux
matching'' \cite{ABEK, Kneipp}, showing that the vortex and monopole
together form a smooth soliton complex. Below we elaborate this
monopole-vortex complex in more detail.  

\subsection{Generalities}

The first crucial observation is that such a smooth complex requires
a compatible orientation between that of the vortex (in the
color-flavor mixed space) and that of the monopoles (in the color
space). For concreteness and for example the $(1,1)$ corner of
$SU(N)_{C+F}$ is selected for the winding for the vortex solution
Eq.~(\ref{LEvortexSU2}), and the associated monopole solution is
accordingly embedded in the $SU(2)$ subgroup of the color $SU(N+1)$
group, in the $(1, N+1)$ plane. Naturally, both the monopole and
vortex can be rotated into any other directions by color-flavor
transformations; in order to maintain the energy they must be rotated
simultaneously (see below).

The relative sign of the flux in Eqs.~(\ref{VortexFlux}) and
(\ref{Monoflux}) indicates that the monopole studied in the
preceding subsection is actually positioned at the rightmost endpoint
of the vortex (as in Fig.~\ref{MattiaandColor}). The outgoing magnetic flux
(in the direction of $S^{3}$) of the monopole is carried away in the
vortex bundle on its left. 

Another crucial observation is about the behavior of the gauge
field. The monopole field in the ``singular'' gauge,
Eq.~(\ref{Manogauge}), exhibits the well-known Dirac string
singularity along the negative $z$ direction, Eq.~(\ref{Dirac2}).
This behavior matches precisely that of the gauge field along the
vortex core, Eq.~(\ref{Dirac1}), confirming further the relative
positioning of the vortex with respect to the monopole.   
The Dirac string singularity Eqs.~(\ref{Dirac1}),(\ref{Dirac2}) is
entirely harmless as it appears multiplied by the squark field in the
squark kinetic term $|{\cal D}_\mu q|^2$; the latter drops to zero along
the vortex core. The color magnetic field is smooth everywhere, as
noted already.

Note that the choice of the particular gauge (the so-called singular
gauge) needed for matching the Dirac singularity of the gauge field,
is precisely the one which guarantees the smooth field configuration
everywhere. In particular, in this gauge no fields ``wind'' around
either the vortex axis or the monopoles. The scalar fields approach
their vacuum expectation value everywhere outside the complex, as in
Eqs.~(\ref{qsing}) and (\ref{NAmonop}). See Fig.~\ref{ComplexFunc}. 
 
In order to describe the field configurations interpolating the vortex
and monopole, $\{q, A, \Phi\}^{\rm MV}$, we make an Ansatz of the form
\begin{align}
q &= 
\begin{pmatrix}
q_1(\rho, z) & \\
& q_2(\rho, z) \mathbf{1}_{N-1}
\end{pmatrix} \ ;  \non
A_{\rho} & = \frac{\cos\theta}{r} 
  (S_{2} \cos\varphi - S_{1} \sin\varphi) \, \Delta(\rho, z) \ ; \non 
A_{\varphi}  &= \frac{1}{\rho}  \left[
\frac{f(\rho, z)}{N}
\begin{pmatrix}
1 \\
& \mathbf{1}_N & \\
& & -N
\end{pmatrix}
+ \frac{f_{\rm NA}(\rho, z)}{N} 
\begin{pmatrix}
N-1 & & \\
& -{\mathbf 1}_{N-1} & \\
& & 0
\end{pmatrix} 
- \sin\theta (S_{1} \cos\varphi + S_{2} \sin\varphi)\, \Delta(\rho, z)
\right]\non
A_{z} &= - \frac{\sin\theta}{r} 
(S_{2} \cos\varphi - S_{1} \sin\varphi) \, \Delta(\rho, z) \ ; \non 
\phi &= \left(v_{1} + \frac{\lambda(\rho, z)}{\sqrt{2N(N+1)}}\right)
\begin{pmatrix}
1 \\
& \mathbf{1}_N &  \\ 
& & -N
\end{pmatrix} 
+ \frac{\lambda_{\rm NA}(\rho, z)}{\sqrt{2N(N-1)}}
\begin{pmatrix}
N-1 &  & \\
& -{\mathbf 1}_{N-1} & \\
& & 0
\end{pmatrix} \ . \label{complex}
\end{align} 
The profile functions $\{q_{1}, q_{2}, f, f_{\rm NA}, \Delta,
\lambda, \lambda_{\rm NA}\}$, with appropriate boundary conditions, can
be determined numerically as we will do in the following section. 

\subsection{Numerical solution}\label{Numerical}

In order to study these configurations numerically, we note that if
the $SU(N)$ and $U(1)$ coupling constants are set to be equal our
monopole-vortex complex is exactly the monopole-vortex complex
generated by the symmetry breaking  
\beq 
SU(2) \stackrel{v_{1}}{\longrightarrow} U(1) 
  \stackrel{v_{2}}{\longrightarrow} \mathbf{1} \ , \qquad 
v_{1} \gg  v_{2} \ , \label{simplestBisyt}
\eeq
\emph{embedded} in a larger color-flavor space. It is therefore
sufficient for our purpose here to consider the minimal case,
Eq.~(\ref{simplestBisyt}). 
The generators of the $SU(2)$ group are $t^a = \tau^a/2$, where
$\tau^a$ are the Pauli matrices and the scalar field has the
form: 
\[  \phi^a = -\sqrt{2}\, m \, \delta^{a3} + \lambda^a\ , \]
where $\lambda^a$ is the fluctuation around the VEV.
The Lagrangian is then: 
\begin{align}
\mathcal{L} = -\frac{1}{4g^2} (F_{\mu\nu}^a)^2 
+ \frac{1}{g^2} |\D_\mu\phi^a|^2 + |\D_\mu q^i|^2
- \frac{g^2}{8} \left|-\xi \delta^{a3} + \nu\lambda^a 
  + q_i^\dag \tau^a q^i \right|^2 
- \left|\left[ m\mathbf{1}_2 - m \tau^3 
  + \frac{1}{\sqrt{2}} \lambda^a \tau^a \right] q^i \right|^2 ,
\end{align}
where we have set 
\beq
\xi \equiv 4 \,\mu \,m \ , \qquad
\nu \equiv 2\, \sqrt{2}\,\mu \ , \nonumber
\eeq
and our convention for the covariant derivatives and field strength 
is
\begin{align}
F_{\mu\nu}^a &= \de_\mu A_\nu^a - \de_\nu A_\mu^a 
  - \epsilon^{abc} A_\mu^b A_\nu^c \ ; \\
\D_\mu q &= \de_\mu q + \frac{i}{2} A_\mu^a \tau^a \, q \ ; \\
\D_\mu\phi^a &= \de_\mu \phi^a - \epsilon^{abc} A_\mu^b \phi^c\ .
\end{align}
After the symmetry breaking at $v_{1}$,  the second color component of 
the squark field becomes massive, so we can set 
\[ q = \begin{pmatrix} q_1(r,z) \\ 0 \end{pmatrix}\ . \]
The equations of motion for the system are:
\begin{align}
\D_\mu F^{\mu\nu a} &= \epsilon^{abc} \left[ 
\phi^{\dag b} \D^\nu \phi^c 
  + \phi^b \left(\D^\nu \phi^c\right)^\dag\right] 
+ \frac{i g^2}{2} \left[ q_i^\dag \tau^a \D^\nu q^i
  - \left(\D^\nu q_i \right)^\dag \tau^a q^i \right] \ , 
\label{gaugecompleta} \\
\D^\mu \D_\mu \phi^a &= - \frac{\nu g^4}{8} \left( -\xi
\delta^{a3} + \nu \lambda^a + q_i^\dag \tau^a q^i \right) 
 - \frac{g^2}{\sqrt{2}} q_i^\dag \tau^a \left( m\mathbf{1}_2 - m \tau^3 
  + \frac{1 }{ \sqrt{2}} \lambda^b \tau^b \right) q^i \ , 
\label{scalarecompleta} \\  
\D^\mu \D_\mu q &= -\frac{g^2}{4} \left(-\xi\delta^{a3} 
+ \nu\,{\rm Re}(\lambda^a) + 
\Tr\,(q^\dag\tau^a q)
\right)(\tau^a q) 
-\left|m\mathbf{1}_2 - m
\tau^3 + \frac{1}{\sqrt{2}}\lambda^a\tau^a\right|^2 q
\ , \label{quarkcompleta} 
\end{align}
where we have defined $|X|^2=X^\dag X$. 
In order to solve these equations numerically, we introduce an Ansatz
adequate for the $SU(2)$ theory, which is somewhat simpler than
Eq.~(\ref{complex}):  
\begin{align}
A_\rho &= \frac{z}{\rho^2 + z^2} \, 
  (\tau_2 \cos{\varphi} - \tau_1 \sin{\varphi}) \, 
  \frac{f(\rho, z)-1}{2} \ ;  \non
A_z &= \frac{\rho }{ \rho^2 + z^2 } \, 
  (\tau_1 \sin{\varphi} - \tau_2 \cos{\varphi}) \, 
  \frac{f(\rho, z)-1}{2}\ ;   \non
A_\varphi &= -\frac{1 }{ \sqrt{\rho^2 + z^2}} \, 
  (\tau_1 \cos{\varphi} + \tau_2 \sin{\varphi}) \, 
  \frac{f(\rho, z)-1}{2} + \tau_3 \, 
  \frac{1 }{ 2 \rho} \ell(\rho, z)\ ; \non
\phi^a &= -\sqrt{2} \, m \, \delta^{a3} + \lambda^a \ , \qquad  
\lambda^{a}= \delta^{a 3}  \,  s (\rho, z)\ ;  \non
q &= \begin{pmatrix} q_1(r,z) \\ 0 \end{pmatrix}\ . \label{MVCmplx} 
\end{align}
Substituting this Ansatz into the equations of motion, written
extensively in Eqs.~(\ref{gaugeeq})-(\ref{quarkeq}), one obtains 
the coupled differential equations
Eqs.~(\ref{profgauge1})-(\ref{profquark}). 
A priori this system of equations seems to be an overdetermined system
with respect to the function $f(\rho,z)$. However as we have shown in
the previous sections, the chosen Ansatz is well suited for both the
monopole and the vortex, and hence we have assumed the existence of a
solution to all the equations, solving only the system
(\ref{profgauge3},\ref{profgauge4},\ref{profscalar},\ref{profquark}). After
the solution was obtained, we plugged it into the constraint equations
(\ref{eq:constraint_scalars}-\ref{eq:constraint_squarks}) as well as
the remaining second order equation \eqref{profgauge2} and indeed
verified that the solution satisfies all equations. 

In solving the system of these differential equations the relaxation
method is very useful. We introduce a fictitious time dependence into
each of the profile functions and then write all the equations as:  
\beq
E_i = \frac{\de h_i }{ \de t}
\eeq
where $E_i$ denotes the equation of motion for the profile function
$h_i$, which is obeyed when $E_i = 0$. It is important that the
equation $E_i$ is of second order in spatial derivatives and 
that the sign of the Laplace operator is positive: 
$E_i = \de_j^2 h_i + \cdots$ with $j$ summing over spatial
dimensions. In this way the equation with the fictitious
time-dependence resembles (a modified form of) the well-known heat
equation, and when a stationary solution has been found, 
i.e.~$\de_t h_i = 0$ the equation of motion for $h_i$ has been
obtained. 

\begin{figure}[htp]
\begin{center}
\includegraphics[width=0.7\linewidth]{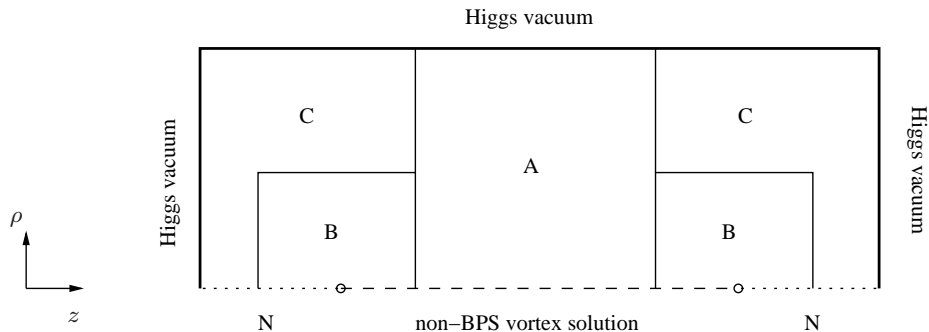}
\caption{The figure shows how the initial conditions are made by
  patching approximate solutions together. A denotes the non-BPS
  vortex solution, B the BPS monopole solution and C is the Higgs
  vacuum. The boundary conditions are also shown, the thick line
  denotes the Higgs vacuum, the dashed denotes the vortex solution
  while the dotted line denotes Neumann boundary conditions. }
\label{fig:bic}
\end{center}
\end{figure}

\begin{figure}[htbp]
\begin{center}
\includegraphics[width=0.5\linewidth]{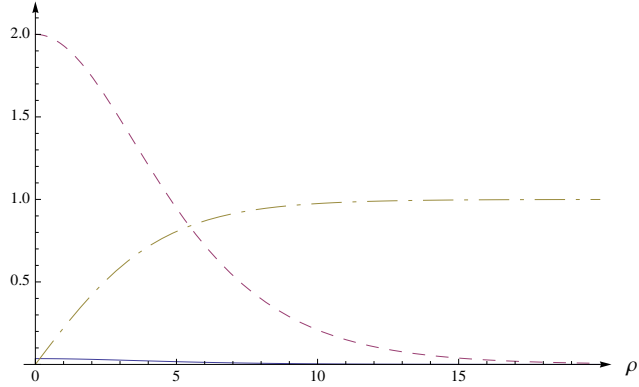}
\caption{The profile functions in the transverse plane in the vortex
  region (far from the monopole). The solid line denotes the adjoint
  scalar field $\frac{1}{\sqrt{2}}s(\rho)$, the dash-dotted line
  denotes the squark field $q(\rho)$ and finally dashed line denotes
  the gauge field function $\ell(\rho)$. The values of the parameters
  in this figure has been chosen as: 
  $g=1,m=2,\nu=2\sqrt{2}\mu=0.1,\xi=\sqrt{2}/5\sim 0.28$.
} 
\label{VortFuncIm}
\end{center}
\end{figure}

First we need to impose some reasonable initial conditions (i.e.~a
guess of the right solution) for the profile functions at $t=0$ and
let the system evolve till a static solution is found. This was done
by patching together three different solutions to the equations of
motion.
The first solution is the non-BPS solution of the vortex (of infinite
length), which is easily obtained numerically, see
Fig.~\ref{VortFuncIm} and also \cite{AEV}. 
The second solution needed is for the monopole and here it turned out
to be most efficient to simply use the analytical solution of the BPS
't Hooft-Polyakov monopole. The third solution is simply the Higgs
vacuum. Now patching the solutions together by
using a piecewise function in the $z,\rho$ coordinates, we obtained an
initial ``solution'', see Fig.~\ref{fig:bic}.

The final ingredient in the solving the vortex-monopole complex is
probably the most crucial, namely the boundary conditions. Physical
arguments tell us that the complex is not really semi-classically
stable as it stands. Hence, one could worry that the relaxation method
would shrink the system to merely the Higgs vacuum. In order to
circumvent that, we implemented the boundary conditions for the system
very similarly to what we did for Abelian soliton junctions in
Ref.~\cite{Bolognesi:2006pp}, viz.~fixing the monopole positions and
hence the string length. In this way, the complex is a minimum of the 
energy -- given the condition we impose. In Fig.~\ref{fig:bic} we have
also shown the boundary conditions (BCs). The three sides far from the
complex all have Dirichlet BCs, i.e.~the Higgs vacuum. At the vortex
string in the middle of the $z$-axis also Dirichlet BCs are imposed,
namely the non-BPS vortex solution described above is used as the
boundary condition -- this is what fixes the string length. For the
remaining pieces on the $z$-axis, we have imposed Neumann BCs due to
cylindrical symmetry of the problem. 
 
\begin{figure}[htbp]
\begin{center}
\mbox{\subfigure[gauge profile function $f(\rho,z)$]{\includegraphics[width=0.48\linewidth]{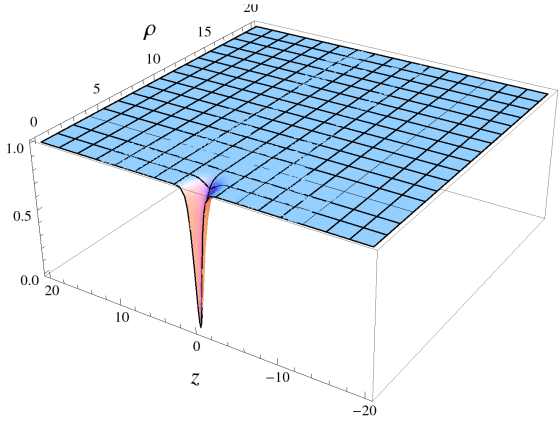}}~
\subfigure[gauge profile function $\ell(\rho,z)$]{\includegraphics[width=0.48\linewidth]{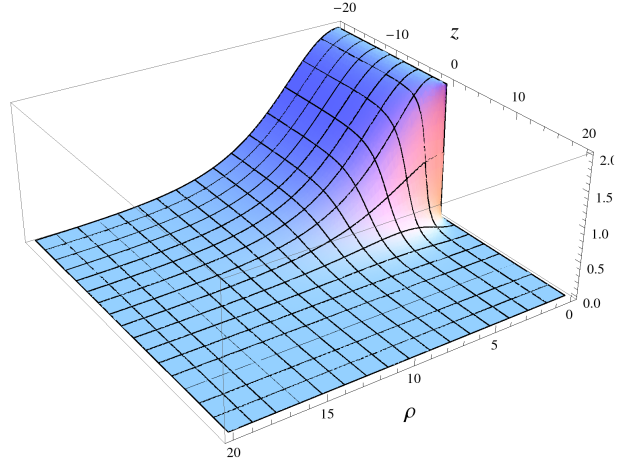}}}\\
\mbox{\subfigure[scalar profile function $\frac{1}{\sqrt{2}}s(\rho,z)$]{\includegraphics[width=0.48\linewidth]{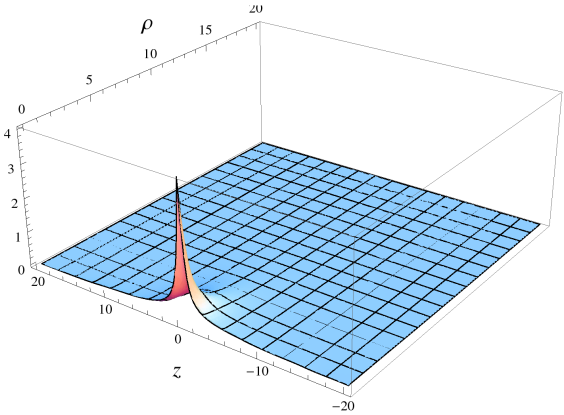}}~ 
\subfigure[squark profile function $q(\rho,z)$]{\includegraphics[width=0.48\linewidth]{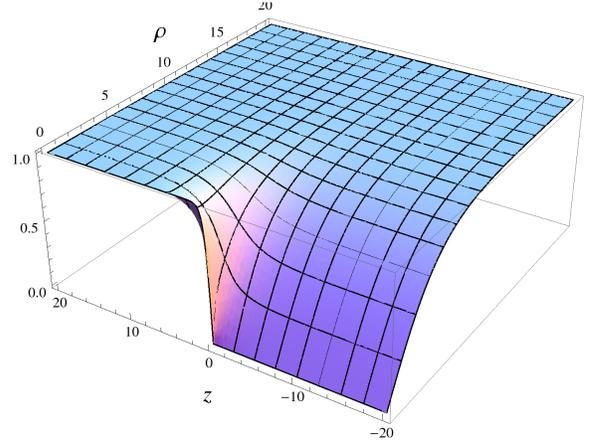}}}
\caption{The behavior of the four profile functions for the gauge and
  scalar fields of Eq.~(\ref{MVCmplx}). Note that the gauge and squark
  fields approach quickly the familiar vortex behavior: the gauge
  field is strongest along the vortex core where the squark field
  drops to zero. Note also that between the figures for $\ell(\rho,z)$ 
  and for $q(\rho,z)$ the $z$ axis is inverted in order to exhibit
  better their non-trivial ($\rho, z)$ dependence. Away from the
  monopole-vortex complex all scalar fields reach quickly and
  uniformly their vacuum expectation values. The values of the parameters
  in this figure has been chosen as: 
  $g=1,m=2,\nu=2\sqrt{2}\mu=0.1,\xi=\sqrt{2}/5\sim 0.28$. }
\label{ComplexFunc}
\end{center}
\end{figure}

The numerical calculation\footnote{The numerical analysis has been
  done with the Mathematica package.} gave the results shown in 
Fig.~\ref{ComplexFunc} for the profile functions $f(\rho,z)$,
$l(\rho,z)$, $s(\rho,z)$, and $q(\rho,z)$. Note that both scalar
fields (the adjoint and fundamental) approach their VEVs smoothly far
from the monopole-vortex complex in all directions. The
monopole-vortex complex is immersed in the Higgs vacuum. 
Far from the monopole centers, the fields reduce to those of the
(non-BPS) vortex, Fig.~\ref{VortFuncIm}, whereas they reduce to the 't
Hooft-Polyakov monopole solution in the small spherical region near
the monopole center.

\begin{figure}[htbp]
\begin{center}
\mbox{\subfigure[]{\includegraphics[width=0.48\linewidth]{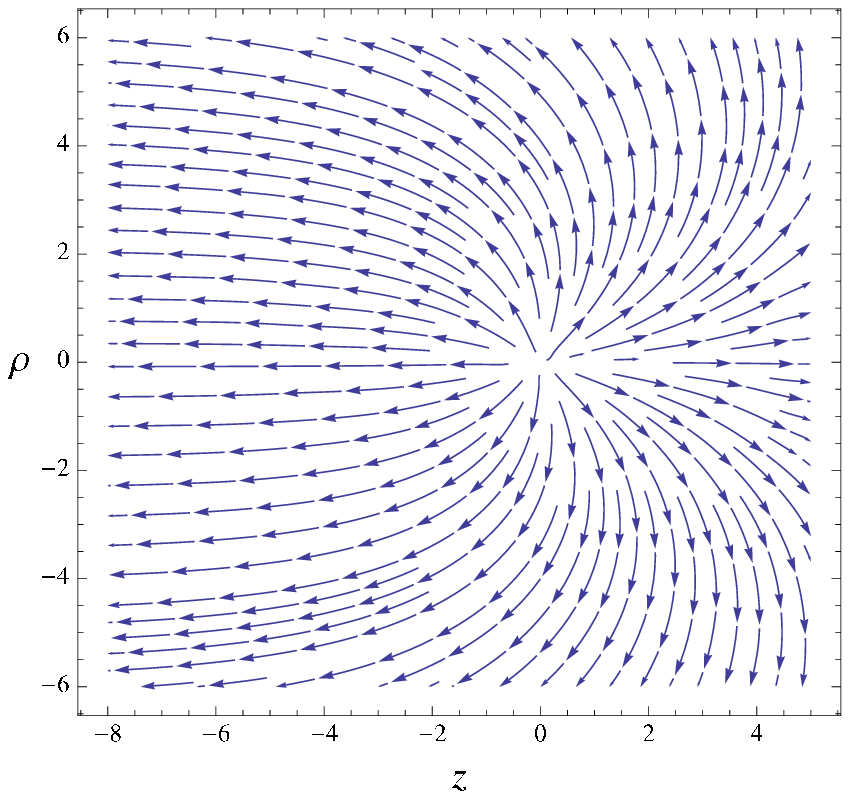}}~~ 
\subfigure[]{\includegraphics[width=0.48\linewidth]{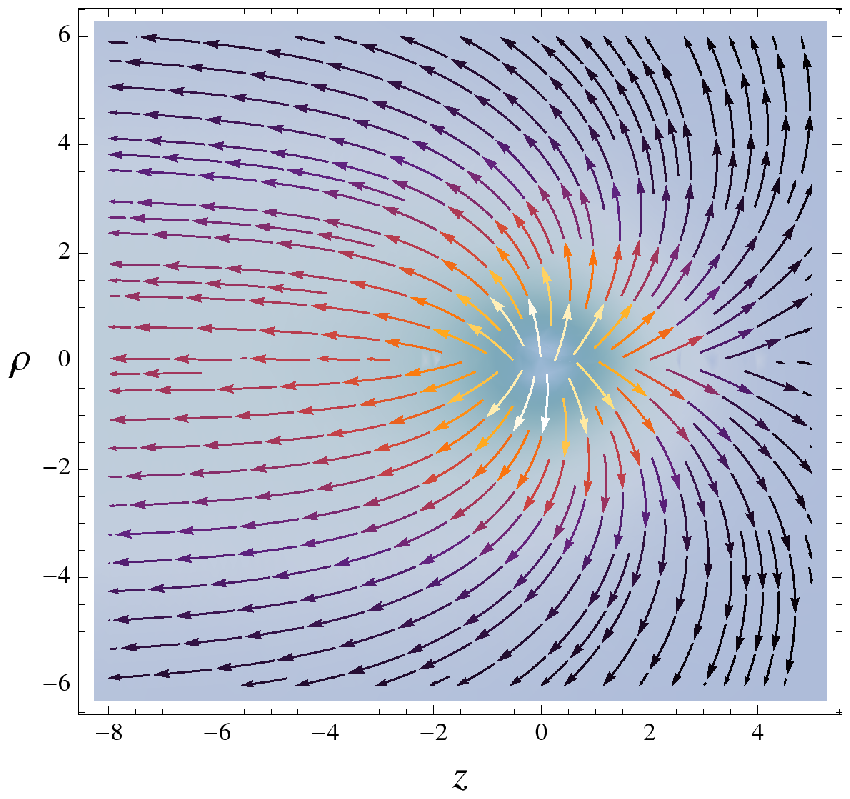}}}
\caption{The behavior of the magnetic field in the soliton complex
  near the monopole region. In (a) is shown a stream line plot while
  the intensity of the magnetic field is also shown in (b) by means of
  the color scheme. For negative values of
  the cylindrically radial coordinate $\rho$ the plot is simply a
  mirror, i.e.~in order to illustrate a cross section of the
  system. The values of the parameters 
  in this figure has been chosen as: 
  $g=1,m=2,\nu=2\sqrt{2}\mu=0.1,\xi=\sqrt{2}/5\sim 0.28$.} 
\label{MattiaandColor}
\end{center}
\end{figure}

\begin{figure}[htbp]
\begin{center}
\includegraphics[width=0.8\linewidth]{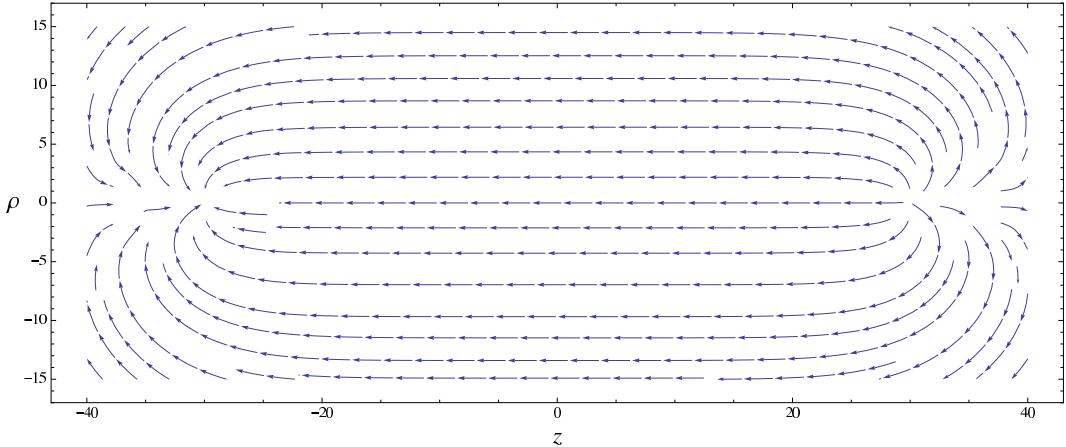}
\caption{The magnetic field in the complete
  monopole-vortex-antimonopole soliton complex. For negative values of
  the cylindrically radial coordinate $\rho$ the plot is simply a
  mirror, i.e.~in order to illustrate a cross section of the system. The values of the parameters
  in this figure has been chosen as: 
  $g=1,m=2,\nu=2\sqrt{2}\mu=0.1,\xi=\sqrt{2}/5\sim 0.28$. }
\label{MattiaMVA}
\end{center}
\end{figure}

These profile functions give rise to the magnetic field shown in 
Figs.~\ref{MattiaandColor} and \ref{MattiaMVA}; as expected the
magnetic flux coming out of the monopole (isotropically near the
monopole center and in the direction of $S^{3}$) is carried away in a 
vortex bundle located on its left, and gets reabsorbed by the
antimonopole sitting at the far left extreme of the vortex.  

\subsection{Macroscopic picture and duality}

Recently the monopole-vortex complex in the simplest gauge symmetry
breaking (\ref{simplestBisyt}) 
\beq 
SU(2) \stackrel{v_{1}}{\longrightarrow} U(1) 
  \stackrel{v_{2}}{\longrightarrow} \mathbf{1}\ , \qquad
v_{1} \gg v_{2}\ , \label{simplestBisytagain}
\eeq
has been discussed \cite{Lahiri, KMO} in the London limit, in this 
limit the monopole is a point and the vortex a line, without width.
In this approximation it is possible to perform the electromagnetic
duality transformation explicitly, and the resulting dual theory can
be solved for the electric and magnetic fields explicitly, in the
presence of the monopole-vortex complex. In the presence of a static
heavy monopole of unit charge sitting at ${\bf r}= {\bf 0}$, and
with a semi-infinite vortex extending on its left, the magnetic (and
electric if $\theta \ne 0$) fields are given by \cite{KMO} 
\beq  
E_{i} = F_{0i} = \frac{\theta g^{2}}{8 \pi^{2}} \, B_{i}^{\rm (mon)}
\ , \qquad 
B_{i} = \frac{1}{2} \epsilon_{ijk} F_{jk} = B_{i}^{\rm (mon)} 
  + B^{\rm (vor)} \delta_{i}^{3}\ ,  \label{remark}
\eeq
where 
\beq
B_{i}^{\rm (mon)} = \frac{1}{g}  \de_{i}  G({\bf r}) \ , \qquad 
B^{\rm (vor)}= \frac{m^{2}}{g} \int_{-\infty}^{0} \,
  dz^{\prime} \, G(x, y, z-z^{\prime}) \ ,\label{using}
\eeq
and $G({\bf r})$ is the Green function, having the Yukawa form
\beq 
G({\bf r}) = \frac{4\pi}{-\Delta + m^{2}}\, \delta^{3}({\bf r}) 
 = \frac{e^{-m r}}{r}\ , \qquad 
m \equiv \frac{g \, v_{2}}{\sqrt{2}} \ . \label{using2}
\eeq
Such a construction, in the case of a more general symmetry
breaking 
\[ SU(N+1)  \stackrel{v_{1}}{\longrightarrow}   SU(N)\times
U(1)    \stackrel{v_{2}}{\longrightarrow} \mathbf{1}\;, \]  
$(N\ge 2)$ and in the presence of a color-flavor diagonal symmetry, is
not known; but as noted above, in the approximation $g_{1}=g_{N}$
the results, Eqs.~(\ref{remark})-(\ref{using2}), represent the
color-magnetic or electric flux in the $S_{3}$ direction 
(Eq.~(\ref{SU2})).

\subsection{Orientational zero modes}\label{Orientation}

Let us now return to the symmetry breaking of our interest here
\beq
SU(N+1)_{\rm color} \otimes SU(N)_{\rm flavor} 
  \stackrel{v_{1}}{\longrightarrow} 
U(N)_{\rm color}  \otimes SU(N)_{\rm flavor} 
\stackrel{v_{2}}{\longrightarrow} SU(N)_{C+F} \ , \qquad  
v_{1} \gg v_{2}\ , \label{simplestCF}
\eeq
and consider the monopole-vortex complex (\ref{complex}). The
color-flavor diagonal $SU(N)_{C+F}$  symmetry (\ref{colorflavor})  is
an exact symmetry of our  $SU(N+1)$ system in the Higgs phase: it is
not an approximate symmetry of the low-energy $SU(N) \times U(1)$
gauge system.  
On the other hand, the monopole-vortex complex (\ref{complex}) is not
invariant under the full  $SU(N)_{C+F}$ symmetry: it transforms as  
\beq 
q^{U} &=& 
\begin{pmatrix}
U & \\
& 1
\end{pmatrix}\, q  \, U^{-1} \ , \qquad 
A_{i}^{U} = 
\begin{pmatrix}
U & \\
& 1
\end{pmatrix}\,  A_{i}  \, 
\begin{pmatrix}
U^{-1} & \\
& 1
\end{pmatrix} \ , \qquad 
i=\rho, \phi \ ,  \non
\phi^{U} &=& 
\begin{pmatrix}
U & \\
& 1
\end{pmatrix}\, \phi \, 
\begin{pmatrix}
U^{-1} & \\
& 1
\end{pmatrix}\ , \label{SUNRotationMV}
\eeq
where $U \in SU(N)_{C+F}$. Of course, not all  $SU(N)_{C+F}$ elements
transform the configuration non-trivially. 
The complex (\ref{complex}) is obviously invariant under the $SU(N-1)$
subgroup generated by   
\beq
\begin{pmatrix}
0 & & \\
& T^A & \\
& & 0
\end{pmatrix} , \qquad  
A=1,2,\ldots, (N-1)^{2}-1 \ .
\eeq
The $U(1)$ subgroup of $SU(N)_{C+F}$, generated by 
\beq 
T^{(0)} =
\begin{pmatrix}
N-1 &  \\
& -{\bf 1}_{N-1} 
\end{pmatrix} \ . \label{decomp}
\eeq 
acts in a subtler way on the monopole-vortex complex. The pure vortex
configuration (\ref{LEvortex}) of the low-energy theory is clearly
invariant under such a $U(1)$ transformation, but the monopole-vortex
complex is not obviously invariant.

The problem is that the complex (\ref{complex}) lives in a larger,
$SU(N+1)$ gauge space, i.e.~it has necessarily non-vanishing color
components in the $(N+1)$-th row  and/or $(N+1)$-th column. 
However, decomposing Eq.~(\ref{decomp}) as
\beq 
\left(\begin{array}{cc}T^{(0)} &  \\    & 0\end{array}\right) = T^{(1)} + (N-1) \, S_{3} \ , \qquad
T^{(1)} =
\begin{pmatrix}
\frac{N-1}{2} & &  \\
& -{\bf 1}_{N-1} & \\
&  & \frac{N-1}{2}
\end{pmatrix} \ ,
\eeq
in terms of two, $(N+1) \times (N+1)$ matrices, one sees that
$\{q,A,\phi\}^{\rm MV}$ is invariant under $T^{(1)}$ transformations
whereas they transform under $e^{i \alpha S_{3}}$ simply as 
\beq 
\{q, A, \phi\}^{\rm MV} \to \left.\{q, A, \phi\}^{\rm MV}\right|
\{{S_{1}\to  S_{1} \cos\alpha - S_{2} \sin\alpha; \,  
  S_{2}\to  S_{2} \cos\alpha + S_{1} \sin\alpha}\}\ .  
\eeq
This can be easily seen to be equivalent to the shift of the azimuthal
angle 
\beq 
\varphi \to \varphi - \alpha\ ;
\eeq
in other words, it is equivalent to (i.e.~can be undone by) a spatial
rotation of angle $\alpha$ around the monopole-vortex axis,
$\hat{z}$.

To summarize, non-trivial zero modes (\ref{SUNRotationMV}) of the
complex are generated by 
\beq 
\begin{pmatrix}
0 & {\bf b}^\dag \\
{\bf b} & {\bf 0}_{N-1}
\end{pmatrix} \ , \label{Goldstonelike}
\eeq
where the complex $(N-1)$-component vector ${\bf b}$ represents the
local inhomogeneous coordinates of  
\beq 
\mathbb{C}P^{N-1} \sim  \frac{SU(N)}{SU(N-1)\times U(1)} \ ,
\eeq
just as what was found by studying exclusively the low-energy vortex
solutions (\ref{coset})-(\ref{invariant}).

\subsection{Demise and resurrection of the $SU(N)$ symmetry}

In the small region around the monopole, (\ref{HEapprox}), the tiny
VEV $v_{2}$ may be effectively set to zero; as the squark fields are
then trivial ($q\equiv 0$) the transformations (\ref{SUNRotationMV})
look locally as global color transformations of the monopole
solution. 

This shows that the isomorphism between the vortex moduli
\cite{ABEKY} and the monopole moduli\footnote{Here we neglect the
  $U(1)$ zero modes related to the electric charge of the monopole;
  in the Higgs phase they survive only in the vicinity of the monopole
  center.  See also Ref.~\cite{KMO}.} -- both found to be
$\mathbb{C}P^{N-1}$ -- is certainly not a coincidence.  Nevertheless,
attempts to understand the moduli space of the \emph{monopoles}
(representing $\pi_2(SU(2)/U(1))$) as something arising from its
various possible embeddings in the larger color space  $G/H=
SU(N+1)/SU(N)\times U(1)$, face inevitably the known difficulties: 
\begin{itemize}
  \item[(i)] Topological obstruction \cite{CDyons};
  \item[(ii)] Non-normalizable gauge zero modes \cite{DFHK}; 
  \item[(iii)] Non-local nature of Goddard-Nuyts-Olive (GNO) duality
    \cite{GNO}\;. 
\end{itemize}
The first two issues have been discussed many times in the literature,
and need not be reviewed here (see however below for observations
concerning these points). 
It is perhaps worthwhile however to recall the third point, which is
apparently the simplest of the three and at the same time, the
deepest. The GNO quantization condition \cite{GNO}
\beq
2 \, {\beta \cdot \alpha} \in \mathbb{Z} 
\eeq
where $\alpha$ are non-vanishing root vectors of $H$, and where 
the asymptotic gauge field is written as (in an appropriate gauge) 
\beq
F_{ij} = \epsilon_{ijk} B_k = 
\epsilon_{ijk}\frac{r_k}{r^3}  (\beta\cdot {\bf H}) \ , \qquad 
\beta = \alpha^* \ ,
\eeq
in terms of the Cartan subalgebra generators ${\bf H}$ of $H$, tells
us that the set of degenerate monopoles are labeled by the {\it dual }
weight vector $\beta$.   
The duality implied by such a formula is clearly a natural
generalization of the electromagnetic duality. The transformations of
the monopole under the dual, magnetic group correspond to some
non-local field transformations in the original description\footnote{
  The fact that in most papers in which these issues are discussed the
  gauge group is taken to be $SU(N)$ appears to have helped to
  obscure this point, as the dual of $SU(N)/Z_{N}$ is again $SU(N)$.
  It is however sufficient to generalize the discussion, for instance,
  to the case of a gauge symmetry breaking 
\beq 
\frac{SO(2N+2)}{\mathbb{Z}_{2}}  \to  
\frac{SO(2N)\times U(1)}{\mathbb{Z}_{2}} \ : \label{USp}
\eeq
in this case the monopole transforms according to the dual of
$SO(2N)/\mathbb{Z}_{2}$: under $Spin(2N)$. The monopoles are
predicted to transform according to one of the $2^{N-1}$ dimensional
spinor representations \cite{GJK}.}. 
Simply, it is not the right question to ask how the monopole solutions 
associated with the gauge symmetry breaking $G \to H$
transform under the ``unbroken''  global color subgroup $H$. 

Let us return to the symmetry breaking (\ref{simplestCF})
\beq 
SU(N+1)_{\rm color} \otimes SU(N)_{\rm flavor} 
  \stackrel{v_{1}}{\longrightarrow} 
U(N)_{\rm color}  \otimes  SU(N)_{\rm flavor} 
  \stackrel{v_{2}}{\longrightarrow}  SU(N)_{C+F} \ , \qquad
v_{1} \gg v_{2} \ . \label{simplestCFbis}
\eeq
The ``magnetic monopole'' solution arising from the breaking at the
scale $v_{1}$ cannot be rotated without spending energy under  
the $SU(N)_{\rm color}$:  the latter is broken by the smaller squark
VEV, $v_{2}$, Eq.~(\ref{VEVS}). It costs energy to vary the embedding
of the $SU(2)$ subgroup $S_{i}$ within the $SU(N+1)_{\rm color}$
(Eq.~(\ref{SU2})), as the squark fields have a preferred color-flavor
locked direction (Eq.~(\ref{LEvortexSing})). Forcing so (rotating the
$SU(2)_{\rm color}$ embedding) would distort the complex and yield a
higher-energy configuration (an excited state). 
In other words, the subtle issues of the non-normalizable zero modes
and the topological obstruction for defining the global color (i) and
(ii), have been converted into an explicit breaking of the symmetry. 

However, the system has an exact color-flavor diagonal $SU(N)_{C+F}$
symmetry.  Individual monopole-vortex complex configurations break
it. They live in the continuous moduli of the solution space
\[  SU(N)_{C+F}/ U(N-1)\sim  \mathbb{C}P^{N-1} \ .  \]
The simultaneous color-flavor $SU(N)_{C+F}$ transformations acting on
the original fields keep the energy unchanged; they \emph{induce}
movements on $\mathbb{C}P^{N-1}$.  The complex transforms as in the
fundamental multiplet ${\bf N}$ of an $SU(N)$.  

This then represents a new, exact non-Abelian symmetry for the
monopole: the latter transforms according to the fundamental
representation of an $SU(N)$ group.

The fluctuations of the associated zero modes propagate only along
the vortex;  this makes sense because, according to the standard lore,
the dual gauge group is in a confinement phase, the original local
gauge group being in the Higgs phase.  The monopoles are confined by
the color-magnetic vortex, which in the dual picture represents the
color-electric confining string.

\section{Conclusion} 

In this paper we elaborated further on some earlier analyses
\cite{ABEK,Duality} of the monopole-vortex complex ``soliton'',
arising in systems with hierarchically broken gauge symmetries. In a
vacuum with an exact unbroken color-flavor symmetry (color-flavor
locked phase), such a complex acquires orientational zero modes,
analogous to those extensively studied recently in the context of
non-Abelian vortices \cite{Konishi}-\cite{Eto:2010mu}. The field
configurations are studied both qualitatively and numerically,
verifying that they constitute a smooth ``constrained''  minimum of
the energy, with the monopole and antimonopole positions kept fixed.
The whole complex possesses the orientational degeneracy (arising from
the exact color-flavor symmetry):  it implies a new, exact continuous
symmetry for the monopole, confined by the vortex.  This, we argue,
is the semi-classical origin of the dual gauge symmetry.   Their
fluctuations propagate in the monopole-vortex-antimonopole
worldsheet strip, the monopole and antimonopole acting as the source
or sink for the massless (Nambu-Goldstone like) excitations,
propagating along the vortex in between.  Extending the direct
derivation of the $2d$ vortex worldsheet sigma models on
$\mathbb{C}P^{N-1}$ \cite{ABEKY,Shifman:2004dr,Gorsky:2004ad} (or in Hermitian
symmetric spaces, $SO(2N)/U(N)$, $USp(2N)/U(N)$, etc., depending on
the model considered \cite{GJK}) to the case of monopole-vortex
complex is a somewhat non-trivial task, presently under
investigation. Our microscopic study here should serve as the basic
starting point for such an analysis.    

 %The notion that the dual non-Abelian gauge symmetry appearing in the low-energy effective description of a strongly-coupled gauge theory, 
%involves essentially the global, flavor symmetry in some way or other, seems to be gaining force, corroborated by many examples. 
%Nevertheless, our construction illustrates certain aspects ---

%That the origin of the dual, non-Abelian gauge symmetry (under which monopoles transform as multiplets) involves the global, flavor symmetry
%in an essential manner together with the original gauge interactions.
%But that in order to illustrate it it is necessary to take into account the appropriate phase 

\subsection*{Acknowledgments}

SBG gratefully acknowledges a Golda Meir post-doctoral fellowship. 
DD thanks Coll\`ege de France and Fondation Hugot for the hospitality and support during the revision of this paper.
The authors thank J. Evslin and T. Fujimori for discussions. 

\appendix

\section{BPS monopole and vortex solutions}

Here we will briefly check that the BPS equations for the monopole and
the vortex are compatible with our second-order equations of motion.

\subsection{Monopole}

Let us start with the monopole; the BPS equation reads:
\beq
F_{ij} = \epsilon_{ijk} \D_k \phi \ ,
\label{BPSmonopoleeuqations}
\eeq
while the covariant derivative of the latter can be written as
\beq
\D_i F_{ij} = \epsilon_{ijk} \D_i \D_k \phi 
 = \frac{1}{2} \epsilon_{ijk} [\D_i, \D_k] \phi 
 = \frac{i}{2} \epsilon_{ijk} [F_{ik},\phi]
 = i [\phi,\D_j\phi] \ ,
\eeq
where we have used the BPS equation in the last equality. 
This equation agrees with Eq.~(\ref{gaugecompleta}) in the BPS 
limit, with $\phi^a$ real (taking into account a factor of
$1/2$ for the canonical normalization of the kinetic term) and
setting $q = 0$. 

From the BPS equation (\ref{BPSmonopoleeuqations}) we obtain for the 
adjoint scalar
\beq
\D_k \D_k \phi = \frac{1}{2} \epsilon_{ijk} \D_k F_{ij} = 0 \ , 
\eeq
due to the Bianchi identity. This equation agrees with
Eq.~(\ref{scalarecompleta}) in the BPS limit; taking  $\mu = 0$ and
$q=0$. 

\subsection{Vortex}

The vortex BPS equations are for $SU(N)\times U(1)$
\begin{align}
F^0_{ij} &\mp \epsilon_{ij} 
g_0^2 \left[\Tr(q^\dag t^0 q) - \xi\right] = 0 
\ ; \label{bogvortex1}\\ 
F^a_{ij} &\mp \epsilon_{ij} g_N^2 \Tr(q^\dag t^a q) = 0 \ ; 
\label{bogvortex2}\\
\D_i q &\pm i \epsilon_{ij} \D_j q = 0 \ ,  \label{bogvortex3}
\end{align}
where the upper (lower) sign gives a BPS vortex (anti-vortex) and
$a=1,\ldots,N^2-1$ is the $SU(N)$ adjoint index (in the $N=1$ case
only the first of the gauge equations appears). 
We consider here a symmetry breaking pattern like 
$SU(N+1) \rightarrow SU(N)\times U(1)\rightarrow\mathbf{1}$ and the
BPS equations come about at low-energy in this system in 
the limit of $\mu\to 0$ while $\xi=-\sqrt{2}\mu\langle\phi^0\rangle$
is kept constant. We will check this explicitly below.

\subsubsection*{General case: $SU(N+1) \rightarrow SU(N) \times U(1)
  \rightarrow \mathbf{1}$} 

The Lagrangian of the $SU(N+1)$ system is
\beq
\mathcal{L} = 
-\frac{1}{4g^2} (F_{\mu\nu}^A)^2 + \frac{1}{g^2} |\D_\mu\phi^A|^2 
+ |\D_\mu q^i|^2 
- g^2 \left|\mu\phi^A + \frac{1}{\sqrt{2}} q_i^\dag t^A q^i\right|^2 
- \left|\left[m\mathbf{1}_{N+1}+\sqrt{2} \phi\right]q^i\right|^2 \ ,
\eeq
which leads to the equations of motion
\begin{align}
\D_\mu F^{\mu\nu A} &= f^{ABC} \left[ 
\phi^{B \dag} \D^\nu \phi^C 
  + \phi^B \left(\D^\nu \phi^C\right)^\dag\right] 
+ i g^2\left[ q_i^\dag t^A \D^\nu q^i
  - \left(\D^\nu q_i \right)^\dag t^A q^i \right] \ , 
\label{gaugecompletaN} \\
\D^\mu \D_\mu \phi^A &= - \frac{\mu g^4}{\sqrt{2}}
\left(\sqrt{2}\mu \phi^A + q_i^\dag t^A q^i \right) 
 - \sqrt{2}g^2 q_i^\dag t^A \left( m\mathbf{1}_{N+1}  
  + \sqrt{2} \phi \right) q^i \ , 
\label{scalarecompletaN} \\  
\D^\mu \D_\mu q &= -g^2\left(\sqrt{2}\mu\,{\rm Re}(\phi^A) + 
\Tr\,(q^\dag t^A q)
\right)(t^A q) 
-\left|m\mathbf{1}_{N+1} 
+\sqrt{2}\phi \right|^2 q
\ , \label{quarkcompletaN} 
\end{align}
where $A=1,\ldots,(N+1)^2-1$ is the $SU(N+1)$ adjoint index and
$|X|^2=X^\dag X$. 
Let us now truncate the theory to the low-energy $SU(N)\times U(1)$
sector, i.e.~we will integrate out the massive components of the
squarks $q$ and also the adjoint and gauge fields corresponding to
broken generators, as they will have a mass of the order of $|v_1|$. 
The vortex solution requires $\phi^\alpha$, $\alpha=0,1,\ldots,N^2-1$
(which is the adjoint field written in terms of the $SU(N)\times U(1)$
group with corresponding index $\alpha$) to be fixed at the VEV 
$\langle\phi^\alpha\rangle = \delta^{\alpha 0} \phi^0$. 
As already mentioned the BPS limit corresponds to taking $\mu
\rightarrow 0$, with the product $\mu \phi^0$ kept finite. 
We have: $\sqrt{2}\langle\phi^\alpha\rangle t^\alpha + m = 0$. 
Hence, in this limit the second equation is trivially satisfied. 

For the first equation we have:
\beq
\D_i F_{ij}^\alpha &= - i g^2\Tr\left[ q^\dag t^\alpha \D_j q
  - (\D_j q)^\dag t^\alpha q \right] \ , 
\label{vortexgaugecompletaN}
\eeq
We can combine the Eqs.~(\ref{bogvortex1}) and (\ref{bogvortex2})
by setting the two coupling constants to be equal; $g_0=g_N=g$ (and
choosing the upper sign) which yields
\beq
F_{ij} - \epsilon_{ij} \frac{g^2}{2} \left[q q^\dag -
  2\xi t^{0} \right] = 0 \ .
\label{BPScombined}
\eeq
Acting with the covariant derivative yields (note that it acts in the
adjoint way on both terms)
\beq
\D_i F_{ij} = \epsilon_{ij} \frac{g^2}{2} 
\left[(\D_i q) q^\dag + q \D_i q^\dag \right]
= -\frac{ig^2}{2}\left[(\D_j q) q^\dag - q \D_j q^\dag \right] \ ,
\eeq
where we have used the BPS equation \eqref{bogvortex3} in the last
step. 
Multiplying by $t^\alpha$ and taking the trace gives us
Eq.~\eqref{vortexgaugecompletaN}. 

The third equation reduces to:
\beq
\D_i\D_i q = g^2 \left(\Tr(q^\dag t^\alpha q) - \xi
\delta^{\alpha 0}\right) (t^\alpha q) 
= \frac{g^2}{2}\left[q q^\dag - 2\xi t^0\right] q \ , 
\label{vortexquarkcompletaN}
\eeq
with $\xi = -\sqrt{2}\mu \phi^0$.
Using the BPS equation \eqref{bogvortex3} we obtain
\beq
\D_i\D_i q = -\frac{i}{2}\epsilon_{ij} [\D_i,\D_j]q 
= F_{12} \, q \ .
\eeq
Now upon insertion of the field strength \eqref{BPScombined} into the
above equation we see that it is exactly that of
Eq.~\eqref{vortexquarkcompletaN}.

\section{Numerical analysis in $SU(2)$ theory}

Writing out the equations of motion
Eqs.~(\ref{gaugecompleta})-(\ref{quarkcompleta}) explicitly in terms of
the fundamental fields gives:
\subsection*{Gauge fields}
\begin{align}
&\de^i\de_i A^{ja} - \de_i \de^j A^{ia} - \epsilon^{abc} \left(
  2A^{ib} \de_iA^{jc} + \de_iA^{ib}A^{jc} - A_i^b\de^j A^{ic}
  \right) +\left(A^{ia}A_i^b A^{jb} - A^{aj} A_i^b A^{ib}
  \right) \label{gaugeeq} \\ 
 &= \epsilon^{abc} \left[ \lambda^{c\dag} \de^j\lambda^b - (\de^j
   \lambda^{c\dag}) \lambda^b - \sqrt{2} m \delta^{b3} \left(
   \de^j\lambda^c + \de^j \lambda^{c\dag} \right)\right] 
- \left( 2
 A^{ja} \lambda^{b\dag} \lambda^b - A^{jb} \left( \lambda^a
 \lambda^{b\dag} + \lambda^b \lambda^{a\dag} \right) \right) \non
&+ \sqrt{2}m \left( \left[ 2 \delta^{c3} \delta^{ba} - \delta^{b3}
   \delta^{ca} - \delta^{a3} \delta^{bc} \right] A^{jb} (\lambda^c +
 \lambda^{c\dag}) \right) - 4m^2 \left( A^{ja} - \delta^{a3} A^{j3}
 \right) \non
&+ \frac{i g^2}{2} \left[ \delta^{a3}\left( q^\dag \de^j
   q - (\de^j q^\dag) q \right) + i A^{ja} q^\dag
   q\right] \ ;    \nonumber
\end{align}

\subsection*{Scalar fields}

\begin{align}
&\de^i\de_i \lambda^a - 2 \epsilon^{abc} A^{ib} \de_i\lambda^c -
 \epsilon^{abc}\de^iA_i^b \lambda^c + A_i^aA^{ib}
 \lambda^b - \lambda^a A^{ib}A^b_i 
- \sqrt{2} m \left[ \epsilon^{ab3} (\de^i A_i^b) + A^{ib}
  A^{ib}\delta^{a3} - A^{i3} A^{ia} \right] \non
&=  -\frac{\nu g^4}{8} \left[
  \left( -\xi + q^\dag q \right) \delta^{a3} + \nu \lambda^a
  \right] - \frac{g^2}{2} q^\dag \left( \delta^{ab} + i
  \epsilon^{ab3} \right) \lambda^b q \ ; 
\end{align}

\subsection*{Squark fields}

\begin{align}\label{quarkeq}
&\de^i\de_i q + \frac{i}{2} (\de^i A_i^3) q 
+ i A_i^{3} \de^i q - \frac{1 }{ 4} A^{ia}A^a_i q = 
-\frac{g^2}{4} \left(-\xi + \nu{\rm Re}(\lambda^3) + q^\dag q\right)q
-\frac{1}{2}\left(|\lambda^3|^2 + |\lambda^1+i\lambda^2|^2\right)q \ ,
\\
&\frac{i}{2}\de^i(A_i^1 + i A_i^2)q + i(A_i^1 + i A_i^2)\de^i q = 
-\frac{\nu g^2}{4}\left({\rm Re}(\lambda^1) + i {\rm
  Re}(\lambda^2)\right) q 
-\frac{1}{\sqrt{2}}\left(\lambda^{1\dag} + i \lambda^{2\dag}\right)
q \non 
&+\frac{1}{2}\left(\lambda^{3\dag}(\lambda^1+i\lambda^2)
-\lambda^3(\lambda^{1\dag} + i \lambda^{2\dag})\right) q \ . 
\end{align}
Inserting the Ansatz \eqref{MVCmplx} into the equation rewritten in
cylindrical coordinates we obtain the system of equations for the
profile functions: 
\begin{subequations}
\begin{multline}\label{profgauge1}
z \de_{z}^2 f + \rho \de_\rho\de_z f 
+ \frac{ \sqrt{\rho^2+z^2} }{ \rho} \left[\left( 1- \ell \right) \de_\rho f 
+ ( f - 1) \de_\rho \ell\right]
- \frac{2\rho z\de_\rho f}{ \rho^2+z^2 } 
+ \frac{ \rho^2-z^2 }{ \rho^2+z^2 }\de_z f  = \\ 
-  \frac{ z^2 }{ \rho^2 } \frac{(f-1)(1-\ell)}{ \sqrt{\rho^2+z^2}} 
+ \frac{g^2}{2} z ( f - 1 ) q^2  
+ \frac{z ( f - 1 )^3}{ \rho^2+z^2} 
+ \frac{z(f-1)(1-\ell)^2}{\rho^2}  
+ 2z (f-1)(s-\sqrt{2}m) ^2 \ ,
\end{multline}
\begin{multline}\label{profgauge2}
\rho \de_{\rho}^2 f + z \de_\rho\de_z f 
+ \frac{ 2z^2 }{ \rho^2+z^2 } \de_\rho f 
- \frac{ \sqrt{\rho^2+z^2} }{ \rho } \left[(f-1) \de_z \ell + \left( 1- \ell \right)\de_z f\right]
- \frac{z }{ \rho} \left(\frac{\rho^2-z^2 }{
  \rho^2+z^2}\right) \de_z f  = \\ 
-  \frac{z(f-1)(1-\ell)}{\rho\sqrt{\rho^2+z^2}} 
+ \frac{g^2}{2} \rho ( f -1 ) q^2 
+ \frac{\rho( f-1 )^3}{ \rho^2+z^2}
+ \frac{1 }{ \rho} ( f-1 ) \left( 1- \ell \right) ^2 + 2\rho  ( f-1 )
(\sqrt{2}m - s)^2  \ , 
\end{multline}
\begin{multline}\label{profgauge5}
\rho\de_\rho\de_z f - z \de_\rho^2 f
+ \de_z f - \frac{z}{\rho} \de_\rho f
+ \frac{2z(f-1)\de_z\ell}{\sqrt{\rho^2+z^2}}
+ \frac{(f-1)(\rho^2-z^2)\de_\rho\ell}{\rho\sqrt{\rho^2+z^2}}
- \frac{z(1-\ell)\de_z f}{\sqrt{\rho^2+z^2}}
+ \frac{(\rho^2+2z^2)(1-\ell)\de_\rho f}{\rho\sqrt{\rho^2+z^2}}
= \\
\frac{z(f-1)(\ell-2)\ell}{\rho^2} \ ,
\end{multline}
\begin{multline}\label{profgauge3}
\de_{\rho}^2 f + \de_{z}^2 f 
+ 2 \frac{z(f-1)\de_\rho \ell}{\rho\sqrt{\rho^2+z^2}} 
- \frac{z(1-\ell) \de_\rho f}{ \rho\sqrt{\rho^2+z^2}} 
- \frac{(\rho^2 - z^2)\de_\rho f}{ \rho ( \rho^2+z^2 ) }
- \frac{2(f-1) \de_z \ell}{\sqrt{\rho^2+z^2}} 
+ \frac{(1-\ell) \de_z f}{ \sqrt{\rho^2+z^2}} 
- \frac{2z\de_z f}{\rho^2+z^2} = \\
\frac{g^2}{2} (f-1) q^2 
+ 2(s-\sqrt{2}m)^2 (f-1) 
- \frac{z(f-1)(1-\ell)}{ \rho^2\sqrt{\rho^2+z^2}}
+ \frac{(f-1)}{\rho^2} + \frac{(f-1)^3}{\rho^2+z^2} \ ,
\end{multline}
\begin{multline}\label{profgauge4}
\de_{\rho}^2 \ell + \de_{z}^2 \ell -\frac {1}{\rho} \de_\rho \ell 
+ \frac{3\rho ( f -1 ) ( \rho \de_z f - z \de_\rho f )}{
  (\rho^2+z^2)^{3/ 2}}  = 
\frac{z(f-1)^2}{ (\rho^2+z^2)^{3 / 2}} -\frac{(1-\ell)(f-1)^2}{
  \rho^2+z^2 } + \frac{g^2 }{ 2} \ell q^2 \ , 
	\end{multline}
\beq\label{profscalar}
\de_{\rho}^2 s + \de_{z}^2 s + \frac{1}{\rho} \de_\rho s = 
\frac{1 }{ 2} g^2 q^2 s 
+ \frac{\nu g^4}{8}(q^2 -\xi + \nu s) 
+ \frac{(s - \sqrt{2}m)(f-1)^2}{ \rho^2+z^2} \ .
\eeq
\beq\label{profquark}
\de_{\rho}^2 q + \de_{z}^2 q + \frac{1 }{ \rho} \de_\rho q =
\frac{1}{2} s^2 q +\frac{g^2}{4} (q^2 - \xi + \nu s)q
+ \frac{(f-1)^2q}{2( \rho^2+z^2 )} + \frac{\ell^2 q}{ 4 \rho^2} \ .
\eeq 
\end{subequations}
Note that plugging Eq.~\eqref{profgauge5} into Eq.~\eqref{profgauge1}
yields Eq.~\eqref{profgauge3}. 

\subsection*{Constraint equations}

\begin{subequations}
\beq
\rho\de_z f - z\de_\rho f 
+ \frac{2(f-1)}{s-\sqrt{2}m}\left[\rho\de_z s - z\de_\rho s\right] = 0
\ , \label{eq:constraint_scalars}
\eeq
\beq
2(f-1)\left[\rho\de_z q - z\de_\rho q\right]
+q\left[\rho\de_z f - z\de_\rho f\right] 
+\frac{1}{\rho}(f-1)(\sqrt{\rho^2+z^2} - z)q = 0 \ . 
\label{eq:constraint_squarks}
\eeq
\end{subequations}

\end{document}